\documentclass[twocolumn,showpacs,preprintnumbers,amsmath,amssymb]{revtex4-1}

\usepackage{graphicx}
\usepackage{dcolumn}
\usepackage{bm}
\usepackage{subfigure}
\usepackage{color}

\begin{document}


\title{Wave function mapping of graphene quantum dots with soft confinement}
\author{D. Subramaniam$^1$}
\author{F. Libisch$^2$}
\author{Y. Li$^3$}
\author{C. Pauly$^1$}
\author{V. Geringer$^1$}
\author{R. Reiter$^2$}
\author{T. Mashoff$^1$}
\author{M. Liebmann$^1$}
\author{J. Burgd\"orfer$^2$}
\author{C. Busse$^4$}
\author{T. Michely$^4$}
\author{R. Mazzarello$^3$}
\author{M. Pratzer$^1$}
\email[] {pratzer@physik.rwth-aachen.de}
\author{M. Morgenstern$^1$}
\affiliation{ $^1$II. Physikalisches Institut B and JARA-FIT, RWTH Aachen University, D-52074 Aachen, Germany\\
$^2$Institute for Theoretical Physics, Vienna University of Technology, A-1040 Vienna, Austria\\
$^3$Institute for Theoretical Solid State Physics and JARA-FIT, RWTH Aachen University, D-52074 Aachen, Germany\\
$^4$II. Physikalisches Institut, Universit\"at zu K\"oln, Z\"ulpicherstr. 77, D-50937 K\"oln, Germany}

\date{\today}

\begin{abstract}

Using low-temperature scanning tunneling spectroscopy, we map the local density of states (LDOS) of graphene quantum dots supported on Ir(111).
Due to a band gap in the projected Ir band structure around the graphene K point, the electronic properties of the QDs are dominantly graphene-like.  Indeed, we compare the results favorably with tight binding calculations on the honeycomb lattice based on parameters derived from density functional theory. We find that the interaction with the substrate near the edge of the island gradually opens a gap in the Dirac cone, which implies soft-wall confinement. Interestingly, this confinement results in highly symmetric wave functions.
Further influences of the substrate are given by the known moir{\'e} potential
and a 10 \% penetration of an Ir surface resonance
into the graphene layer.

\end{abstract}

\maketitle

Graphene has moved in short time from first preparation as a small flake \cite{Geim1} towards possible applications such
as high frequency transistors \cite{Avouris}, supercapacitors \cite{supercapacitor} or touch screens \cite{samsung}.
Another exciting perspective is to use graphene quantum dots (QDs) as
spin qubits \cite{Trauzettel}. The basic prerequisite is a very long spin coherence time \cite{diVinc},
which might exist in graphene \cite{Burkard} due to the absence of hyperfine
coupling in isotopically pure material and the small spin-orbit
coupling \cite{Fabian}. First graphene QDs have been produced and probed by
transport measurements \cite{Ponamerenko,Stampfer}.
However, since graphene provides no natural gap, it is difficult to control the
electron number \cite{Libisch}.
Moreover, the 2D sublattice symmetry makes the QD properties very susceptible to the atomic edge configuration \cite{Trauzettel} unlike conventional QDs. As a result, chaotic Dirac billiards have been predicted \cite{ber87}
and were even claimed to be realized \cite{Ponamerenko, Wurm}, i.e. the wave functions are assumed to be rather disordered.
To achieve improved control of graphene QDs, the QD edges must be well defined and a deeper understanding of the QD properties is mandatory.\\
Direct insight into QD properties is provided by scanning tunneling spectroscopy (STS) which maps out
the squared wave functions of QDs \cite{Berndt2} and, at the same
time, determines the shape of the QD atom by atom.  Using STS, we
probe graphene QDs with well defined zig-zag edges supported on an
Ir(111) surface \cite{Lacovig}. These QDs maintain graphene properties
as the filled part of the graphene Dirac cone lies in the Ir projected
band gap \cite{Petikosic}.  By comparing the measured wave functions
with model calculations, we determine the relationship between
geometry and electronic properties and extract general trends. Most
notably, the soft edge potential provided by the interaction of the QD edges with the
substrate enhances the geometrical symmetry of the wave functions, thus rendering
the QD more regular. The susceptibility of the wave functions to the edge configuration is intimately related to the additional
sublattice symmetry (pseudospin) which makes graphene so special
\cite{Geim}. Also the
moir{\'e} pattern induced by the graphene-Ir lattice mismatch \cite{Petikosic} and the hybridization of graphene with an Ir surface resonance are shown to have an influence on the measured wave functions.


\begin{figure}[b!]
\includegraphics[width=8.5cm]{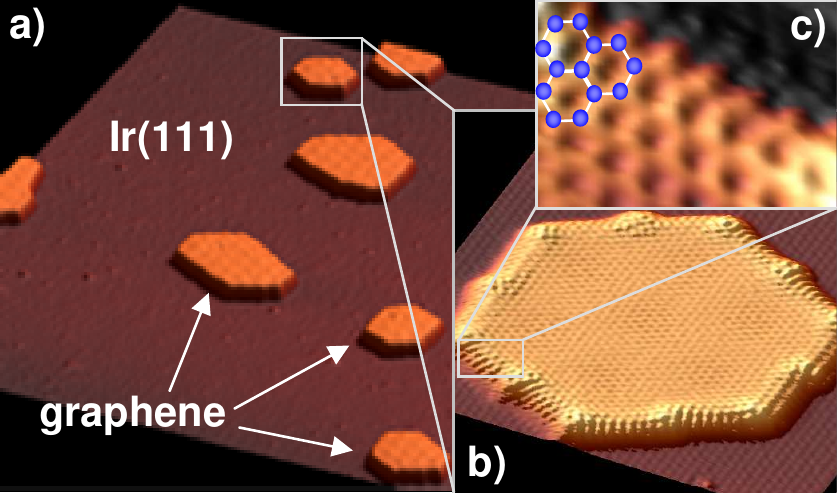}
\caption{\label{morphology}(color online)  (a) ($100\times100$)\,nm$^2$ STM image of Ir(111) covered by monolayer graphene islands; $U=-0.3$~V, $I=0.3$\,nA; (b) atomically resolved ($12\times12$)\,nm$^2$ image of graphene island; (c) magnified view of zigzag edge with graphene lattice overlaid; $U=0.7$\, V, $I=20$\,nA.}
\end{figure}
STM measurements are performed in ultrahigh vacuum at $T=5$\,K \cite{Mashoff1}.
Monolayer graphene islands are prepared by exposing clean Ir(111)
for 4 min to a pressure of 10$^{-5}$ Pa of C$_2$H$_4$ at 300 K and subsequent annealing to 1320\,K (30 s) \cite{NDiyae}. The resulting graphene QDs have diameters of $2-40$\,nm as shown in Fig. \ref{morphology}a. Atomically resolved QD images (Fig. \ref{morphology}b$-$c) reveal the complete enclosure of the QDs by zigzag edges.

\begin{figure}[t!]
\includegraphics[width=8.3cm]{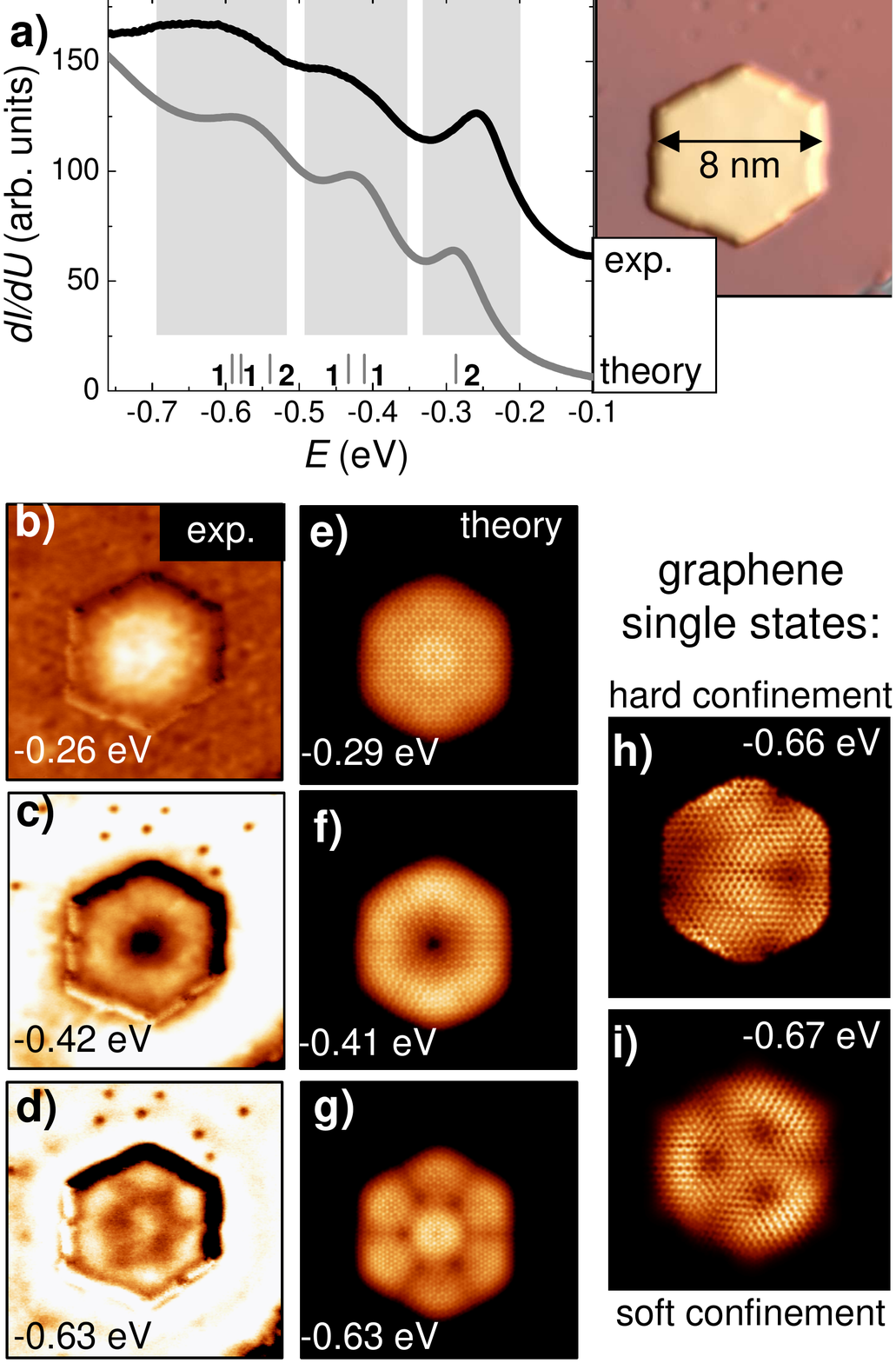}
\caption{\label{confined} (color online) (a) black line: $dI/dU(U)$ curve spatially averaged over the graphene QD shown to the right; $U_{\rm stab}=0.5$\,V, $I_{\rm stab}=0.5$\,nA, $U_{\rm mod}=10$\,mV; grey line: DOS(E) of the same island as obtained by TB calculation (see text); vertical bars mark the calculated eigenstate energies with degeneracies indicated as numbers; (b)-(d) $dI/dU$ images recorded at energies $E=U\cdot e$ as marked; $I=0.2$\,nA; $U_{\rm mod}=10$\,mV. (e)-(g) LDOS maps calculated with soft edge potential at energies indicated; (h),(i): LDOS of an individual state calculated without (h) and with (i) soft edge potential.}
\end{figure}

\begin{figure}[t!]
\includegraphics[width=8.5cm]{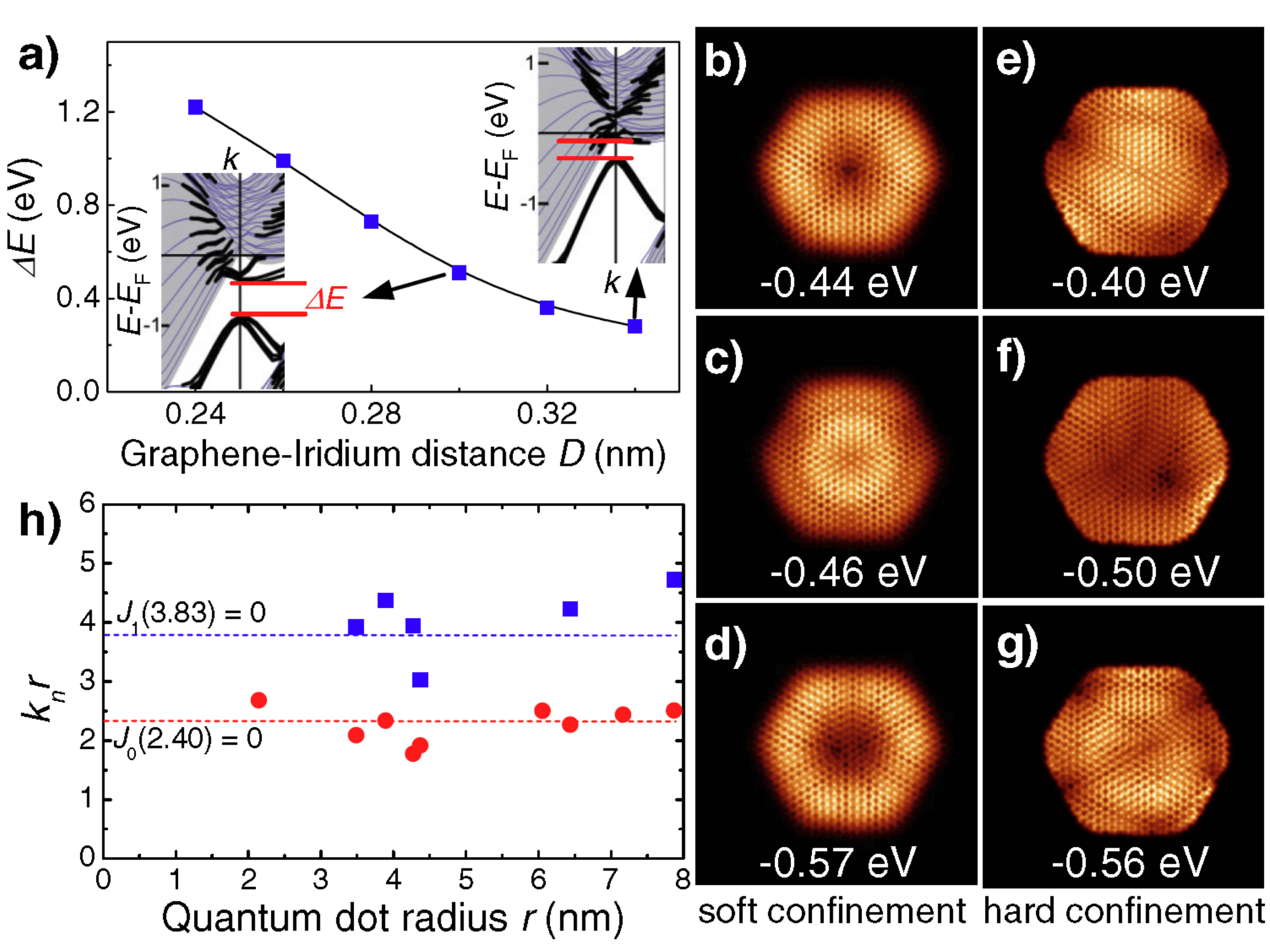}
\caption{\label{theory}(color online) (a) Energy gap $\Delta E$ versus graphene-Ir distance $D$ as deduced from DFT calculations;  insets: band structure around $E_{\rm D}$ for two different $D$ as marked by arrows with $\Delta E$ indicated; grey area: projected bulk bands of Ir; Thick black lines: graphene states; (b)-(g) Calculated LDOS ($=|\Psi|^2$) for individual confined states with energies marked: (b)-(d) with soft edge potential; (e)-(g) without soft edge potential; (h) experimental $k_n \cdot r = E_n/(\hbar v_{\rm D})\cdot r$ for the two
peaks closest to $E_D$ at different average island radius $r$; circles: $n=0$, squares: $n=1$;
dotted lines: zeros of the first two Bessel functions (see text).}
\end{figure}

The local density of states (LDOS) of 15 islands is mapped  by STS. We use a lock-in technique with modulation frequency $\nu=1.4$\,kHz and amplitude $U_{\rm mod}=10$\,mV resulting in an energy resolution $\delta E\approx\sqrt{(3.3\cdot k_BT)^2+(1.8\cdot eU_{\rm mod})^2}=18$\,meV \cite{Morgenstern}. For $dI/dU$ curves, we stabilize the tip at sample voltage $U_{\rm stab}$ and current $I_{\rm stab}$. Figure \ref{confined}a shows a $dI/dU$ curve laterally averaged over the hexagonal QD shown to the right. It displays three maxima below the Dirac point $E_{\rm D}$, which is slightly above the Fermi level $E_{\rm F}$ \cite{Petikosic}. Thus, the peaks belong to confined hole states. Fig.\ref{confined}b-d show $dI/dU$ maps at the peak energies. For the first peak ($U=-0.26$\,V), one maximum of the LDOS in the center of the island appears, a ring shaped structure is observed at $U=-0.42$\,V, and,  a maximum-minimum-maximum sequence from the center towards the rim with an additional star-shaped angular dependence is visible at $U=-0.63$\,V.
We checked that no other LDOS shapes are present at $-1.4 eV \le U \le 0$\,V.
From the sequence of observed LDOS shapes we conclude that they represent confined states of the QD. \\
To model the QD states, we employ third-nearest neighbor tight binding (TB) calculations \cite{TBReich, TBWirtz, Libisch2} using the atomic configuration of the QD found by STM,
\begin{equation}
  H = \sum_{i,s}\left|\phi_{i,s}\right>V_i\left<\phi_{i,s}\right|+
  \sum_{(i,j),s}\gamma_{(i,j)}\left|\phi_{i,s}\right>\left<\phi_{j,s}\right|+h.c.
  \label{H_Graph_TB}.
\end{equation}
The $\gamma_{(i,j)}$ are hopping amplitudes between sites $i$ and $j$
being $\gamma_{(i,j)} = (3.14, 0.042, 0.35)$ eV for
the (first, second, third) nearest-neighbors \cite{TBReich}. The $V_i$
represent local on-site potentials.\\
We first employed a spatially constant $V_i$ within the islands, i.e. hard-wall-confinement. Regular, but also very
 irregular wave functions result, as shown in Fig. \ref{confined}h and Fig. \ref{theory}e-g. The irregular wave functions often display a large intensity at the rim of the QDs and illustrate the sensitivity of graphene QDs to details of the edge configuration \cite{Trauzettel,ber87}. Such irregular shapes, however, were never found in the present STS experiments featuring about 50 different states \cite{supplement}.\\
This failure is related to the two experimental facts that (i) a graphene flake
bends downward from $D=3.4$ ${\rm \AA}$ in the center of a QD to $D=1.6$ ${\rm\AA}$ at its rim
\cite{Lacovig,Atodiresei} and that (ii) the entire graphene flake features a moir{\'e} type corrugation leading to minigaps \cite{Petikosic,Brune,Rader}.

\begin{figure}[t!]
\includegraphics[width=8.5cm]{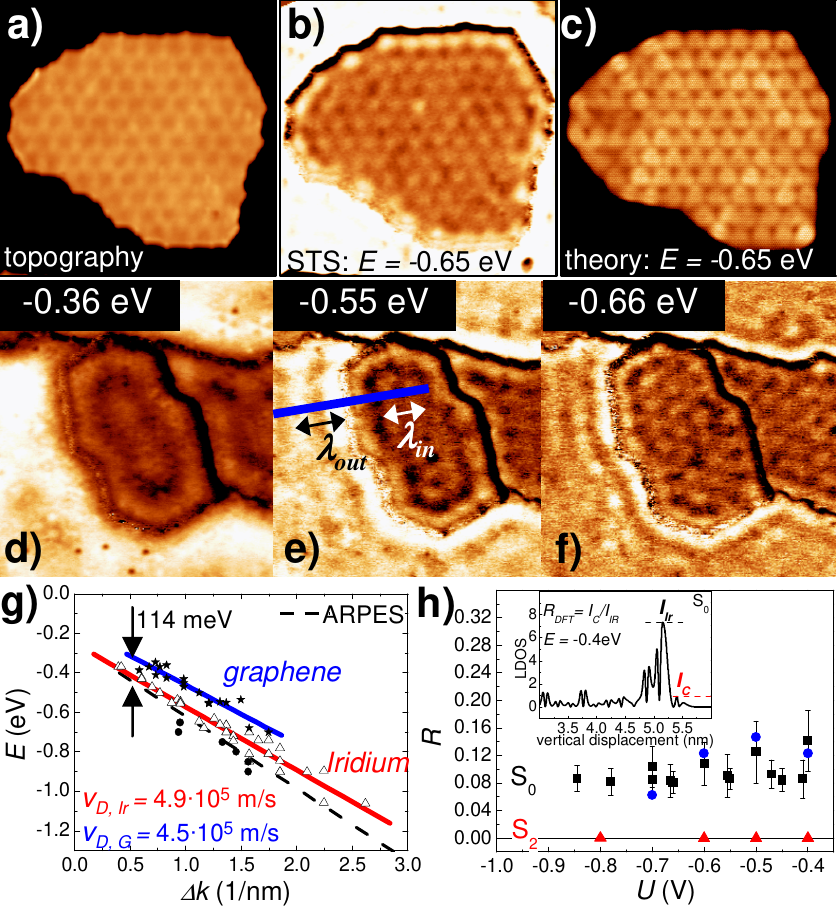}
\caption{\label{dispersion}(color online)
(a) STM image and (b) $dI/dU$ map of a large graphene QD; $30\times30$ nm$^2$, $U=-0.65$\,V, $I=0.5$\,nA, $U_{\rm mod}=10$ mV; (c) calculated LDOS of the same QD at $E=-0.65$\,eV;
(d)-(f) $dI/dU$ maps of a graphene QD recorded at the energies marked; $27\times30$ nm$^2$, $I=0.5$\,nA, $U_{\rm mod}=10$\,mV; deduced wave lengths $\lambda_{\rm out}$ ($\lambda_{\rm in}$) outside (inside) the QD are marked in (e); (g) resulting dispersion relations $E(\Delta k =\pi/\lambda_{\rm in/out})$ inside (stars) and outside (triangles) of the QD as well as from standing waves scattered at Ir(111) step edges (circles); full lines are linear fits with resulting $v_{\rm D}$ indicated; energy offset is marked; dashed line is deduced from photoemission on clean Ir(111) {\protect \cite{Rader}}; (h) relative intensity $R$ of S$_0$ and S$_2$ in graphene as deduced from STS data (squares) and from DFT calculations (S$_0$: circles, S$_2$: triangles); inset: calculated LDOS of S$_0$ at $E=-0.4$ eV along the direction perpendicular to the surface; $I_{\rm Ir}$ and $I_{\rm C}$ as used for determination of $R$ are marked.
}
\end{figure}

To incorporate effect (i) we determined the band structure of graphene by {\it ab initio}
density functional theory (DFT) calculations \cite{QE, supplement} for different graphene-Ir surface distances $D$. Upper and lower limits for $D$ were set by the known
distance between extended graphene layers and Ir$(111)$, $D= 3.4$ ${\rm \AA}$ \cite{Atodiresei} and  the smallest distance found at the edge of a graphene island,
$D= 1.6$ ${\rm \AA}$ \cite{Lacovig}. A proper description of Ir$(111)$ surface states requires thick slabs which makes it unfeasible to use the large $10\times 10$ supercell
necessary to account for the graphene-Ir lattice mismatch.
Therefore, a slightly compressed Ir lattice is used making graphene and Ir(111) commensurate.
This allows us to work with a slab of $24$ Ir layers with graphene on both sides and a vacuum space of $20$ ${\rm \AA}$ between slabs. The insets in Fig. \ref{theory}a exhibit the resulting band structures for two different fixed $D$. The size of the gap $\Delta E_{\rm D}$ is plotted in Fig. \ref{theory}a.
We incorporate the effect of the $D$ dependent band-gap on $V_i$ within the TB through \cite{ber87}:
\begin{equation}\label{eq:pot}
V_{i,\rm{rim}} = \Delta E[D(r_i)]/2 \cdot \sigma_z\,,
\end{equation}
where the Pauli matrix $\sigma_z$ acts on the sublattice degree of
freedom. A homogeneous $V_{i,\rm{rim}}$ would open a gap of size $\Delta E$ at $E_{\rm D}$.
The functional form of $\Delta E [D] = (0.7*(3.6-D[{\rm \AA}])^2+0.23$) eV is taken from the fit to the DFT calculations (Fig. \ref{theory}a). We model the global height variation of a graphene QD by linear increase of $D(r)$ from the rim towards 10 ${\rm \AA}$ inside the island as suggested by the DFT calculations of \cite{Lacovig}. We checked that reasonable modifications do not change
the results significantly \cite{supplement}.\\
To incorporate effect (ii), we added a moir{\'e} potential $V_{i,\rm{m}}$ to $V_i$. Based on the experimentally observed minigap of 200 meV
 \cite{Petikosic,Brune,Rader}, we use a harmonic variation of $V_{i,\rm{m}}$ in each of the three dense packed directions of graphene with a total amplitude of 400 meV \cite{supplement}. Finally, the peak width $\Gamma$ of the eigenstates is adapted to the experiment leading to
$\Gamma(E) =0.33 \cdot \left| E \right|$.
\\
%
The resulting LDOS curve (grey line,
Fig.~\ref{confined}a) as well as the calculated LDOS maps (Fig. \ref{confined}e-g) exhibit excellent agreement with
the experimental data.
%
Importantly, the calculations yield only states that reflect the hexagonal symmetry of the QD shape in agreement with experiment, but none of the irregular states found without smooth confinement \cite{supplement}. This can be rationalized by the suppressed interaction of the confined states with the zig-zag edges, which would break sublattice symmetry \cite{Nakada}. The increased geometrical symmetry is illustrated in Fig. \ref{theory}b-g comparing wave functions of the same quantum dot with soft (hard) confinement leading to symmetric (irregular) states. Thus, softly opening a band gap at the QD edge leads to strongly improved control on the states residing in its interior.\\
To illustrate this crucial finding, we show that the state energies in our QDs can be correctly estimated by a simplified circular flake geometry.
We obtain $E_n=\hbar v_{\rm D} k_n$ with Dirac velocity $v_{\rm D}=10^6$ m/s and $k_n$ deduced from the Bessel functions:
%
\begin{equation}\label{Eq:Bessels}
J_n(k_n\cdot r) = 0,\quad n=0,1,....
\end{equation}
Up to an island area of $A=150$ nm$^2$ (average radius: $r=\sqrt{A/\pi}$), the estimate fits the experimental peak energies to within $\sim 20$ \% for the two lowest energy states (Fig. \ref{theory}h). Larger islands do not follow this trend because of their strong deviation from a circular shape (e.g. Fig. \ref{dispersion}a).
Obviously, neither the sensitive sublattice symmetry of graphene \cite{Trauzettel}, nor the influence of the iridium substrate enter Eq.~(\ref{Eq:Bessels}) showing the simplicity of softly confined graphene QDs. Note, in addition, that the agreement in Fig. \ref{theory}h only uses the peak energies as an experimental reference and does not refer to the
measured LDOS shapes. Thus, peak energies are compatible with $v_{\rm D}=1\pm 0.1 \cdot10^6$ m/s.\\
%
%
%
In larger islands, we observe the influence of $V_{i,\rm{m}}$ on wave function patterns directly, at energies $E < -0.6$\,eV.
Figure \ref{dispersion}a shows an STM topography of a large QD exhibiting a regular moir{\'e} pattern \cite{NDiyae}. The $dI/dU$ map in Fig. \ref{dispersion}b and the calculated LDOS in Fig. \ref{dispersion}c reproduce the moir{\'e} topography albeit with inverted amplitude. The same result is found for all larger islands \cite{supplement}. We checked that normalizing the $dI/dU$ images
to account for a spatially varying tip-surface distance \cite{Wittneven} did not change the LDOS patterns.\\
%
%
One feature, already visible by comparing Fig. \ref{dispersion}b and c, is not accounted for by a spatially varying $V_i$:
a bright rim of the island in the $dI/dU$ image. This rim is found for all islands, but cannot be reproduced by the TB calculations \cite{supplement}. Closer to $E_{\rm D}$, this feature develops into a standing wave pattern
that finds its counterpart outside the island with slightly larger wave length $\lambda$ (Fig. \ref{dispersion}d$-$f). The dispersion relations $E(\Delta k = \pi/\lambda)$ \cite{DK} inside and outside the islands are evaluated as displayed in Fig. \ref{dispersion}e and in \cite{supplement} for 11 islands. They are shown together with results from standing waves at step edges of Ir(111) in Fig. \ref{dispersion}g. The $E(\Delta k)$ curves are linear according to $E=-\hbar v_{\rm D} \Delta k+E_{\rm D}$ with $v_{\rm D}\simeq 4.9 \cdot10^5$\,m/s, $E_{\rm D}=-0.3$ eV outside the island and $v_{\rm D}\simeq 4.5 \cdot10^5$\,m/s, $E_{\rm D}=-0.2$ eV inside the island.
These values agree with those of the Ir surface resonance S$_0$ around $\overline{\Gamma}$ found by photoemission (dashed line) including the energy offset between the two $E(\Delta k)$ curves \cite{Rader}.
The values disagree with $v_{\rm D}$ for the graphene Dirac cone on Ir(111) by a factor of two and with $E_{\rm D}$ for the
Ir S2 surface state by 0.5 eV \cite{Petikosic}. Thus, the standing wave patterns within the QD are attributed to an intrusion of S$_0$ into graphene. The amplitude of the standing wave in the islands $A_{\rm G}$ is found to be close to the amplitude outside the island
$A_{\rm Ir}$ for several islands and energies \cite{supplement}. This is surprising considering the fact that the tip is 0.23 nm further away from the Ir surface, when positioned above graphene, which would suggest a reduction in $dI/dU$ intensity by a factor of 100 \cite{Morgenstern}. However, DFT calculations reveal that S$_0$, exhibiting sp-symmetry, penetrates into graphene. The ratio between the LDOS in the graphene layer $I_{\rm C}$ and the LDOS in the Ir surface layer $I_{\rm Ir}$ is $R_{\rm DFT}= I_{\rm C}/I_{\rm Ir} \simeq 8-12$ \% (inset of Fig. \ref{dispersion}h). For comparison, S$_2$ shows only $R_{\rm DFT}\simeq 0.02$ \%. Figure \ref{dispersion}h favorably compares $R_{\rm DFT}$ of S$_0$ with the data from STS $R_{\rm STS}$ where the apparent $A_{\rm G}/A_{\rm Ir}$ is rescaled according to $R_{\rm STS}= A_{\rm G}/A_{\rm Ir}\cdot e^{\alpha \delta}$ \cite{Morgenstern} with $\alpha = 1.1-1.2/{\rm \AA}$ deduced from $I(z)$ curves and $\delta= 1.1$ $\rm \AA$ being the difference between real height (3.4 $\rm \AA$ \cite{Atodiresei}) and apparent STM height (2.3 $\rm \AA$) of the graphene above the Ir(111). Thus, we can quantitatively reproduce the strength of S$_0$ intrusion into graphene.
A simple explanation for the strong S$_0$ intrusion is not obvious, but we note that, according to DFT, also the d$_z^2$-like surface state S$_1$, located at $E_F$ and exhibiting no dispersion \cite{Petikosic}, pe\-ne\-trates 
into graphene with $R \simeq 10-40$ \% and the $\pi$-electrons of graphene penetrate back into Ir with $R\simeq 1-4$ \%.\\
Finally, we would like to comment on the fact that the S$_0$ state partly dominates the LDOS patterns, while the peak energies are reproduced nicely by the Dirac cone of graphene.
We assume that the life time of the graphene states is large enough to lead to confinement resonances appearing as peaks, while the life time of the S$_0$ is significantly shorter leading only to exponentially decaying standing waves at the step edges of the graphene islands. Indeed, we do not observe peaks within the spectroscopy of the islands, where the standing wave of S$_0$ is dominating the LDOS pattern. Moreover, the standing wave gets always significantly weaker in intensity away from the step edge. Of course, S$_0$ probably influences the LDOS patterns of the small islands as well, which might explain the remaining deviations
between theory and experiment in Fig. \ref{confined} (e.g. (d) and (g)). This subtle interplay between graphene electrons and S$_0$ electrons within the graphene island might also explain the too low $v_{\rm D}=6\cdot 10^5$ m/s
resulting from the analysis of the LDOS pattern of graphene quantum dot states on Ir(111) in ref. \cite{Liljeroth}.  
 
In conclusion, we mapped the LDOS of graphene QDs supported on Ir(111). For small islands, properties of an isolated graphene QD with soft edge potential reproduce the measured wave functions. Most importantly, the soft edge induced by the substrate is required for the experimentally observed high symmetry of the wave functions. Larger islands show an additional standing wave pattern caused by an intruding Ir surface resonance and signatures of the moir{\'e} potential.

We acknowledge helpful discussions with N. Atodiresei, C. Stampfer, G. Burkard, S. Runte, and a referee, as well as financial support by DFG (LI 1050/2-1, MO 858/8-2, BU 2197/2-1), Fonds National de la Recherche
(Luxembourg), and FWF (SFB-F41 VICOM). Numerical calculations are performed on the Vienna
Scientific Cluster (VSC)\\
Note added in proof: During the referee process, two publications  with similar
experimental results have been published \cite{Liljeroth}, which were submitted later than our manuscript.

%

\end{document}



\title{Supplementary information: Wave function mapping in graphene quantum dots}
\author{D. Subramaniam$^1$}
\author{F. Libisch$^2$}
\author{C. Pauly$^1$}
\author{V. Geringer$^1$}
\author{T. Mashoff$^1$}
\author{M. Liebmann$^1$}
\author{R. Reiter$^2$}
\author{Y. Li$^3$}

\author{J. Burgd\"orfer$^2$}
\author{C. Busse$^4$}
\author{T. Michely$^4$}
\author{R. Mazzarello$^3$}
\author{M. Pratzer$^1$}
\email[] {pratzer@physik.rwth-aachen.de}
\author{M. Morgenstern$^1$}
\affiliation{ $^1$II. Physikalisches Institut B and JARA-FIT, RWTH Aachen University, D-52074 Aachen, Germany\\ $^2$ Institute for Theoretical Physics, Vienna University of Technology, A-1040 Vienna, Austria\\ $^3$Institute for Theoretical Solid State Physics and JARA-FIT, RWTH Aachen University, D-52074 Aachen, Germany\\ $^4$II. Physikalisches Institut, Universit\"at zu K\"oln, Z\"ulpicherstr. 77, D-50937 K\"oln, Germany}

\date{\today}


\maketitle

\section{Image gallery of investigated quantum dots}

\begin{figure}[h!]
\includegraphics[width=8.5cm]{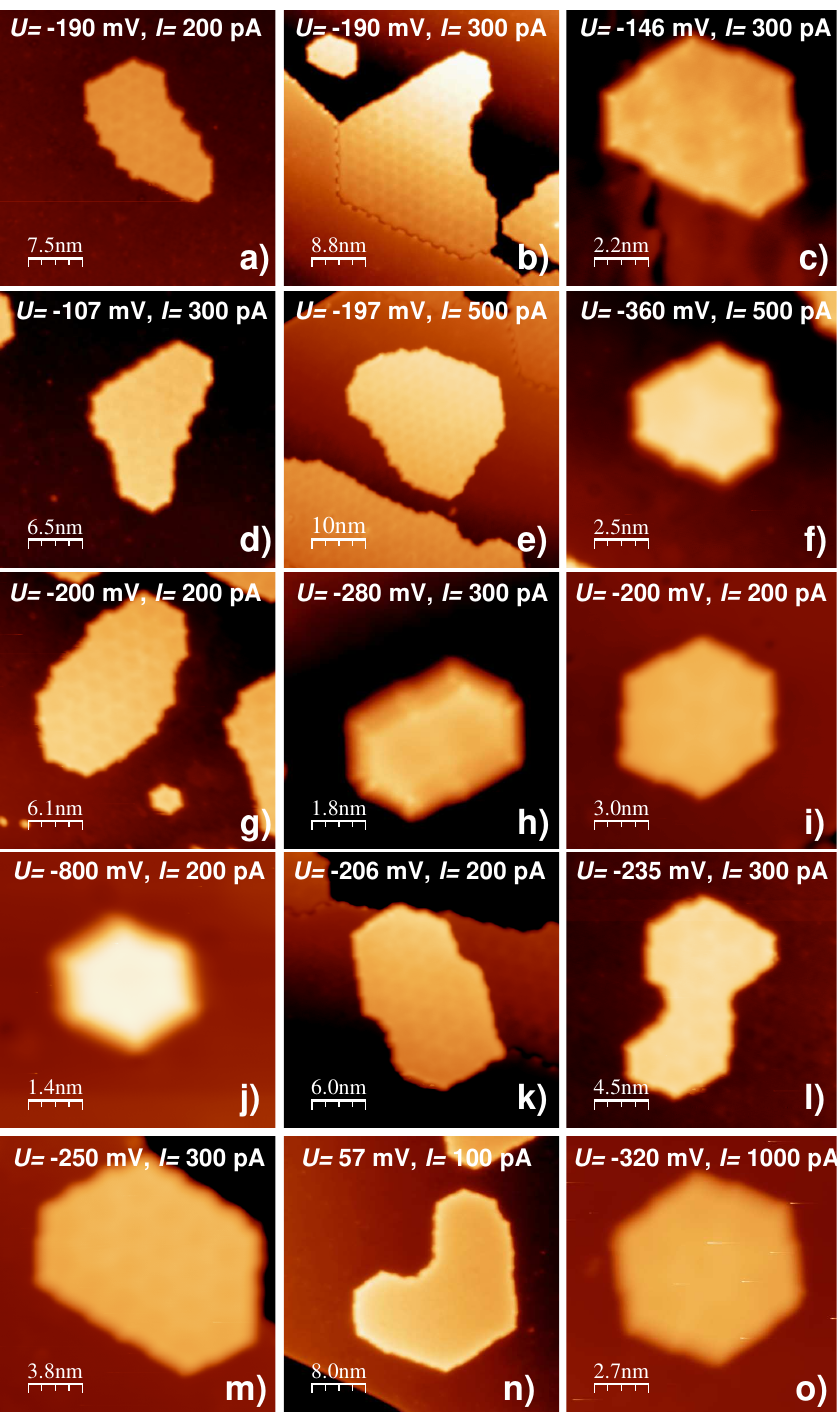}
\caption{\label{overview} Constant current STM images of investigated graphene QDs. Tunneling parameters are marked in the images.}
\end{figure}

Figure \ref{overview} shows an STM image gallery of all investigated graphene quantum dots (QDs) on the Ir(111) substrate. Detailed comparisons of $dI/dU$ images and $dI/dU$ curves with the results from third nearest neighbor tight binding calculations have been performed for theses islands, but only representative data are shown.

\newpage

\section{Density Functional Theory (DFT) calculations}

{\it Ab initio} DFT simulations were carried out using the plane-wave PWSCF code included in the QUANTUM-ESPRESSO package~\cite{QE}.

We firstly studied the relationship between the size of the band gap in graphene and
the graphene-Ir surface distance $D$, ranging from $D= 3.4$ $\rm \AA$ to $D= 1.6$ $\rm \AA$.
A $1\times 1$ cell with the graphene cell parameter was used, which corresponds to a slightly compressed Ir$(111)$ surface.
We use
a slab of $24$ Ir layers with graphene on both sides and a vacuum space
of $20$ ${\rm \AA}$ between slabs.
We employed gradient-corrected exchange correlation functionals~\cite{pbe} and
fully-relativistic ultrasoft pseudopotentials including spin-orbit interactions~\cite{dalcorso}.
The wave functions were expanded in plane waves with a kinetic energy
cutoff of 30 Ry and a charge-density cutoff of 300 Ry.
$20 \times 20 \times 1$ Monkhorst-Pack meshes~\cite{MP} of $k$-points were used for the integration
over the Brillouin zone.

The electronic band structure of the system was computed along the $\bar T$, $\bar T'$ and $\bar \Sigma$ lines of the surface Brillouin zone,
corresponding to the path $\bar \Gamma-\bar K-\bar M-\bar \Gamma$ in the reciprocal space.
The graphene states were identified by projecting the wavefunctions
of the slab on the atomic wavefunctions centered on the C atoms with a threshold of 50 \%.
This allowed us to calculate the gap $\Delta E$ between the lower and upper graphene cone as a function of the graphene-Ir distance (see main text).
The absolute value of the Dirac point energy with respect to the Fermi level $E_{\rm F}$ at equilibrium distance ($D=3.4$ ${\rm \AA}$)
slightly deviates (by 200 meV) from photoemission
experiments \cite{Brune}. The band gap has also been estimated from the photoemission data to be 100 meV \cite{Brune}, i.e.
100 meV smaller that the one found by DFT at the equilibrium distance.
We assume that these discrepancies are caused by the compression of the Ir(111) within the
DFT calculations, but does not affect the general trend of increasing the band gap with shortened graphene-Ir distance.

\begin{figure}[h!]
\includegraphics[angle=0,width=9.0cm]{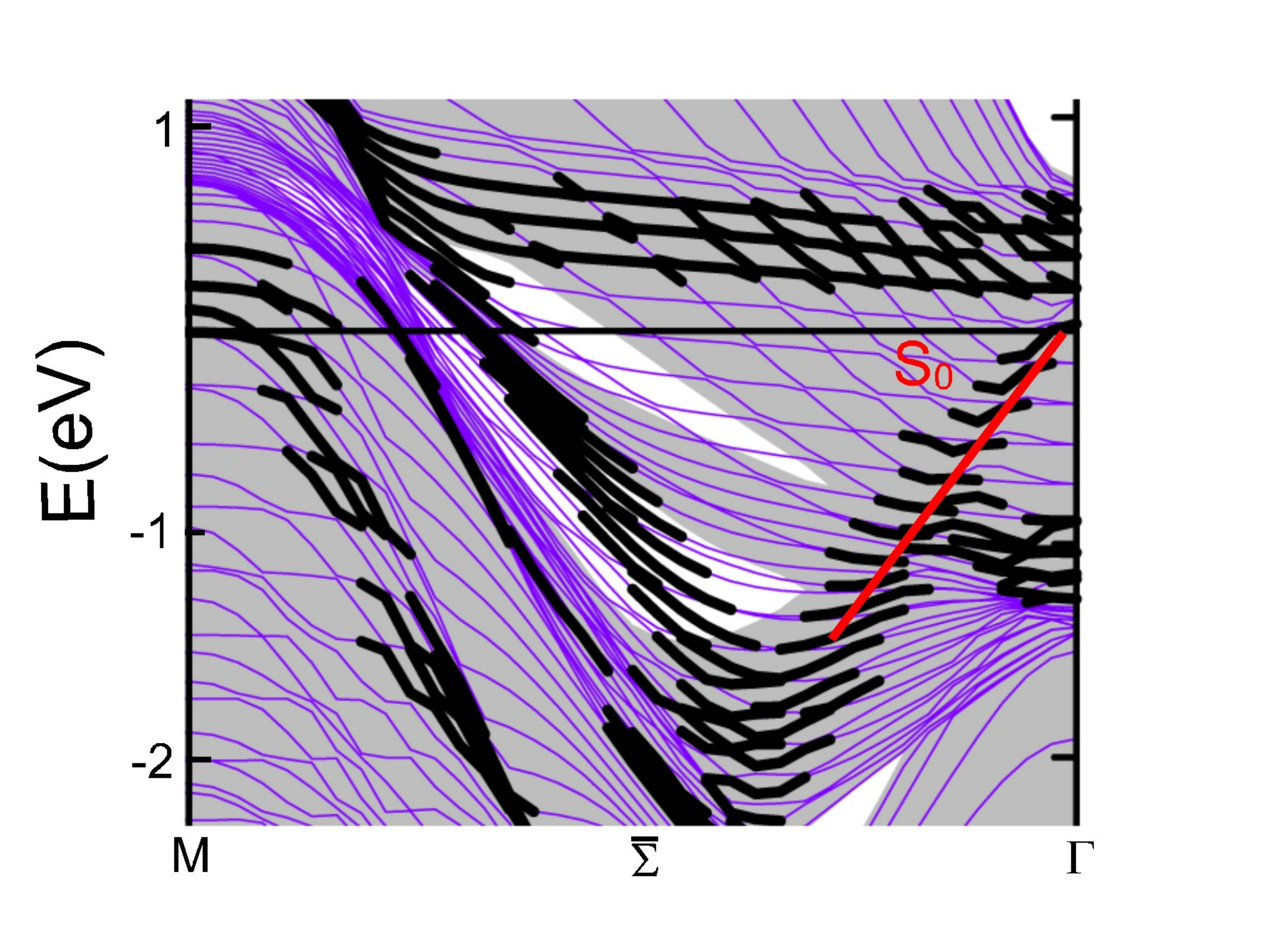}
\caption{\label{band} Band structure of the graphene-Ir system along the $\bar \Gamma-\bar M$ direction. The dispersion of the S$_0$ surface resonance
is indicated by a red line. The state has been identified by requiring that the sum of the squares of the
projections of the state on the orbitals of the atoms
of the surface and subsurface layer be larger than $0.25$.}
\end{figure}

Very recently, Varykhalov {\it et. al.}~\cite{Rader} detected a large Rashba effect on a surface state of Ir$(111)$
(denoted as S$_0$ in the main text) near the $\bar \Gamma$ point: the properties of this state
were found to be hardly affected when the surface is covered with graphene.
We accordingly carried out DFT simulations of both the clean Ir$(111)$ surface with a lattice constant of $a_0=2.758$ $\rm \AA$,
corresponding to the value of $a_{fcc}= 3.90$ $\rm \AA$ for bulk fcc Ir obtained from DFT calculations, and the graphene-Ir system
with $D= 3.4$ $\rm \AA$. For these calculations, we also used a 24-layer Ir slab and identified the Ir$(111)$ surface states
by projecting them onto the atomic orbitals of the surface and
subsurface Ir atoms requiring more than 25 \% of their weight located at these atoms.
We found the surface state S$_0$, which is rather a surface resonance, in both cases.
The band structure of the graphene-Ir system along the
$\bar \Gamma-\bar M$ direction is shown in Fig.~\ref{band} with the surface resonance marked.
The penetration of the surface resonance S$_0$ (as well as of the surface state S$_2$) into the graphene layer was calculated by relating the weight on the C atoms $I_{\rm C}$ to the weight
on the Ir surface atoms $I_{\rm Ir}$ using $R_{\rm DFT}=I_{\rm C}/I_{\rm Ir}$ (see main text).

%
\section{Confined potential used for the calculation}

\begin{figure}[h!]
\includegraphics[width=8.5cm]{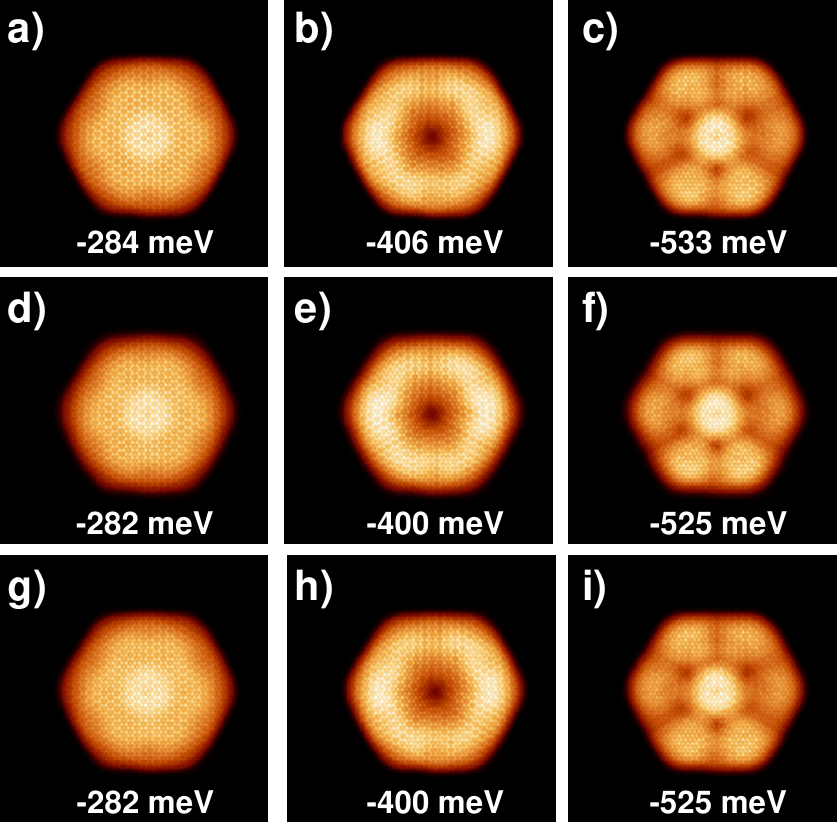}
\caption{\label{fig:edges}{The three lowest-lying eigenstates with
    their eigenenergies marked for different parametrizations of the
    distance development between the graphene quantum dot and the
    iridium substrate at the edge of the quantum dot:
    (a)-(c) %
    linear interpolation $\propto x$ over 15  ${\rm \AA}$, i.e.
    $D(r_\perp) = 3.4 - 0.12 \cdot(15-r_\perp)\cdot\theta(15-r_\perp)$
    with $r_\perp$, $D$ in ${\rm \AA}$ and $\theta(x)$ being the step function;
    (d)-(f) 
    $x^{3/2}$ over 10 ${\rm \AA}$, i.e. $D(r_\perp) = 3.4 - 0.057\cdot(10-r_\perp)^{3/2}\theta\cdot(10-r_\perp)$;
    (g)-(i) %
    quadratic interpolation ($\propto x^2$) over 10 ${\rm \AA}$, i.e. $D(r_\perp) = 3.4-0.018\cdot(10-r_\perp)^2\cdot\theta(10-r_\perp)$;
    in the main manuscript, a linear interpolation over 10 ${\rm \AA}$ was used
    (see main manuscript, Fig.~2).
  }}
\end{figure}

The potential $V_i$ used for the tight binding calculation
is modeled according to the
distance $D$ from the graphene sheet to the Ir surface [see Eq.~(2) in
  main text]. We use a Berry-Mondragon like potential $V_i = \Delta E[D]/2\cdot\sigma_z$
 to model the gap opening at the Dirac point, as taken from our
DFT calculations (see Fig.3a of main text). Consequently, the potential
conserves $K$-$K'$ symmetry \cite{Libisch2} and models the (partial)
hybridization of the carbon $p_z$ orbitals with the Ir states known from our DFT calculations.
Since the graphene quantum dot approaches the Ir surface at the edges, we
obtain a smooth edge confinement that is key to the
formation of states with high symmetry observed in the experimental LDOS
patterns. Near the edges, we assume a linear approach (as the most
simple model) of the graphene sheet towards the Ir substrate from the
equilibrium distance of 3.4 ${\rm \AA}$ to 1.6 ${\rm \AA}$ at the edge over a distance
of 10 ${\rm \AA}$.
To assert that this choice does not influence our conclusions, we have
performed calculations for several different functional forms for
distance between the quantum dot and the Iridium surface
keeping the outmost distance of 1.6  ${\rm \AA}$
fixed. Excluding unphysical, vertical kinks in the shape of the graphene flake, we find no
noticeable changes in the wavefunction patterns (see Fig.~\ref{fig:edges}). The variations in
calculated resonance energies for different types of edge potentials are below 15 meV,
which is smaller than the experimental energy
resolution.\\
Secondly, the lattice mismatch between Ir and graphene is
taken into account by a spatially varying moir{\'e} potential $V_{\rm m}(x,y)$, which consists of a suitable superposition of sinusoidal functions
in accordance with ref. \cite{Atodiresei,NDiyae}
(see main text). In order to obtain the amplitude $V_{\rm max}$ of $V_{\rm m}$,
we perform tight binding calculations of the bandstructure of an infinitely extended graphene sheet in the presence of $V_{\rm m}$ for different $V_{\rm max}$, and extract
the size of the minigap $\Delta$ induced by zone folding \cite{Brune}. Comparing to ARPES measurements, which
feature $\Delta \simeq 200$ meV \cite{Brune}, we obtain $V_{\rm max} = 400$ meV.



\newpage
\section{Calculated states without soft edge potential}
\begin{figure}[h!]
\includegraphics[width=14cm]{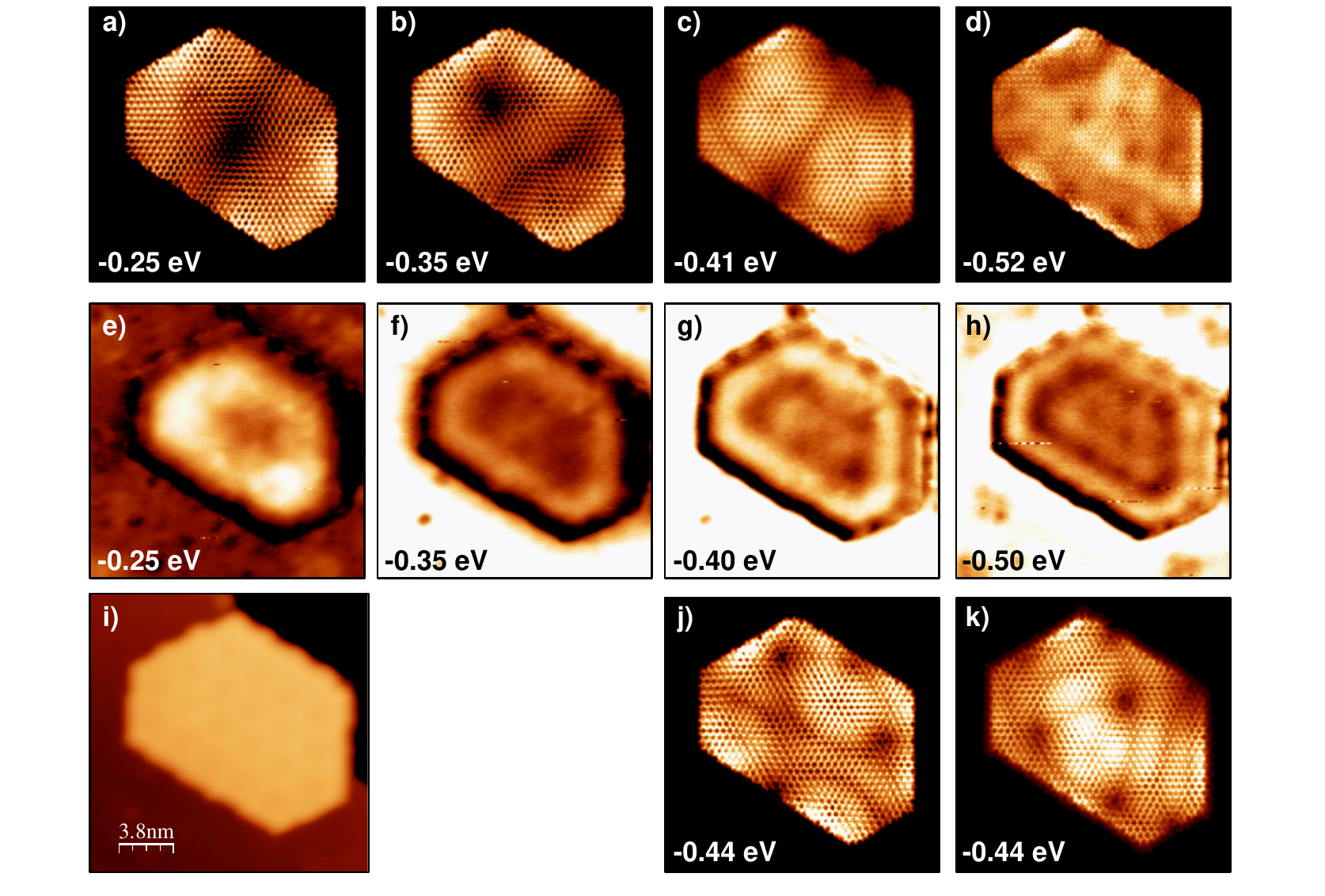}
\caption{\label{without} (a)-(c) calculated squared wave functions of the island shown in (i) at energies marked; (d) calculated LDOS of the same island (consisting of 6 wave functions) at the energy marked; (e)-(h) $dI/dU$ maps of the island shown in (i) at different energies $E=U\cdot e$ as marked; $I=0.3$ nA, $U_{\rm mod}=10$ mV; (i) STM image of the graphene QD; $U=-250$ mV, $I=0.3$ nA;  (j)-(k) calculated squared wave functions
of the QD shown in (i) not found in experiment; all calculations are with abrupt edge.}
\end{figure}

Figure \ref{without}i shows an STM image of a particular island with
the corresponding $dI/dU$ maps displayed in Fig. \ref{without} e$-$h.
We display all patterns of $dI/dU$ maps found in the voltage range between
$U=-0.2$ V and $U= -0.6$ V.  For comparison, wave functions are calculated by
the third nearest neighbor tight-binding calculation neglecting the
soft edge, i.e. $V_i=\infty$ outside the island. Inside
the island, the moir{\'e} potential is maintained.  Superpositions of squared wave
functions at the energies corresponding to the $dI/dU$ images are
shown in Fig. \ref{without} a$-$d. Only the two very similar states belonging to
$K$ and $K'$ are superposed in Fig. \ref{without}a$-$c.
In Figure \ref{without}d already 6 states, which have a very similar energy,
had to be superposed.
The shapes of the wave
functions and the LDOS correspond to the measured $dI/dU$ maps and the
energies reasonably well, but the calculated shapes are less
  regular and more extended towards the edge. Moreover, the
calculated wave functions displayed in Fig. \ref{without} j$-$k do
not resemble the experimental LDOS maps. These wave functions
  feature a strong weight at the edge of the QD. They are suppressed by the soft confinement caused by
   hybridization of the graphene $p_z$
  orbitals with the iridium substrate.

\newpage






\section{States dominated by the {\protect moir{\'e}} potential}

\begin{figure}[h!]
\includegraphics[width=8.5cm]{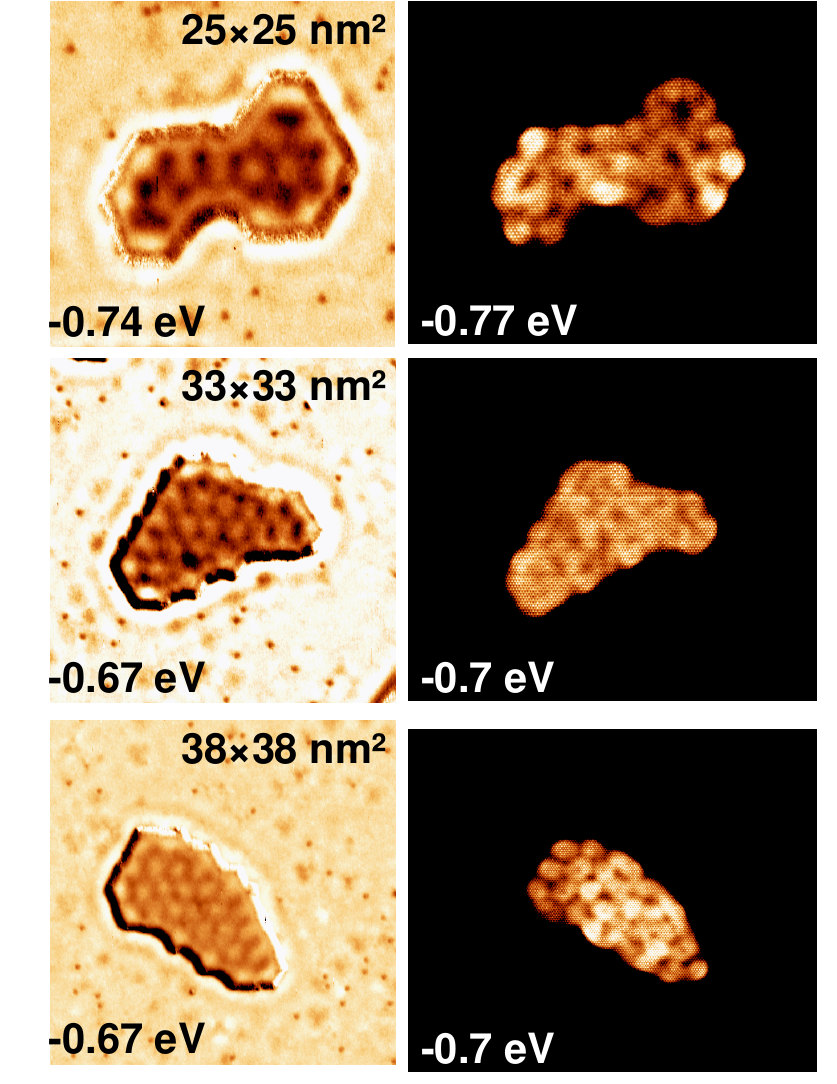}
\caption{\label{moire} $dI/dU$ maps (left images) of different
  graphene QDs imaged at energies below -0.6\,eV ($I=0.4$\,nA,
  $U_{\rm mod} =10$ mV). The images are dominated by the
  moir{\'e} potential and are largely reproduced by the
  calculation (right images).}
\end{figure}

For energies below -0.6\,eV, the $dI/dU$ maps are dominated by the
moir{\'e} potential (see, e.g., the three islands in
Fig. \ref{moire}). The experimental data (left images) correspond to the
calculated LDOS maps (right images) being prone to the moir{\'e}
potential with amplitude 400 meV. Neglecting this moir{\'e}
potential leads to more uniform wave patterns in the LDOS
  maps (not shown). Notice that a bright rim is visible in all
experimental LDOS maps of Fig. \ref{moire} as well as in Fig. \ref{without}(f)-(h), which is not reproduced by the
  calculation and attributed to the Ir surface resonance around $\overline{\Gamma}$ (see main text).

\section{standing waves outside the graphene island}

\begin{figure}[h!]
\includegraphics[width=10cm]{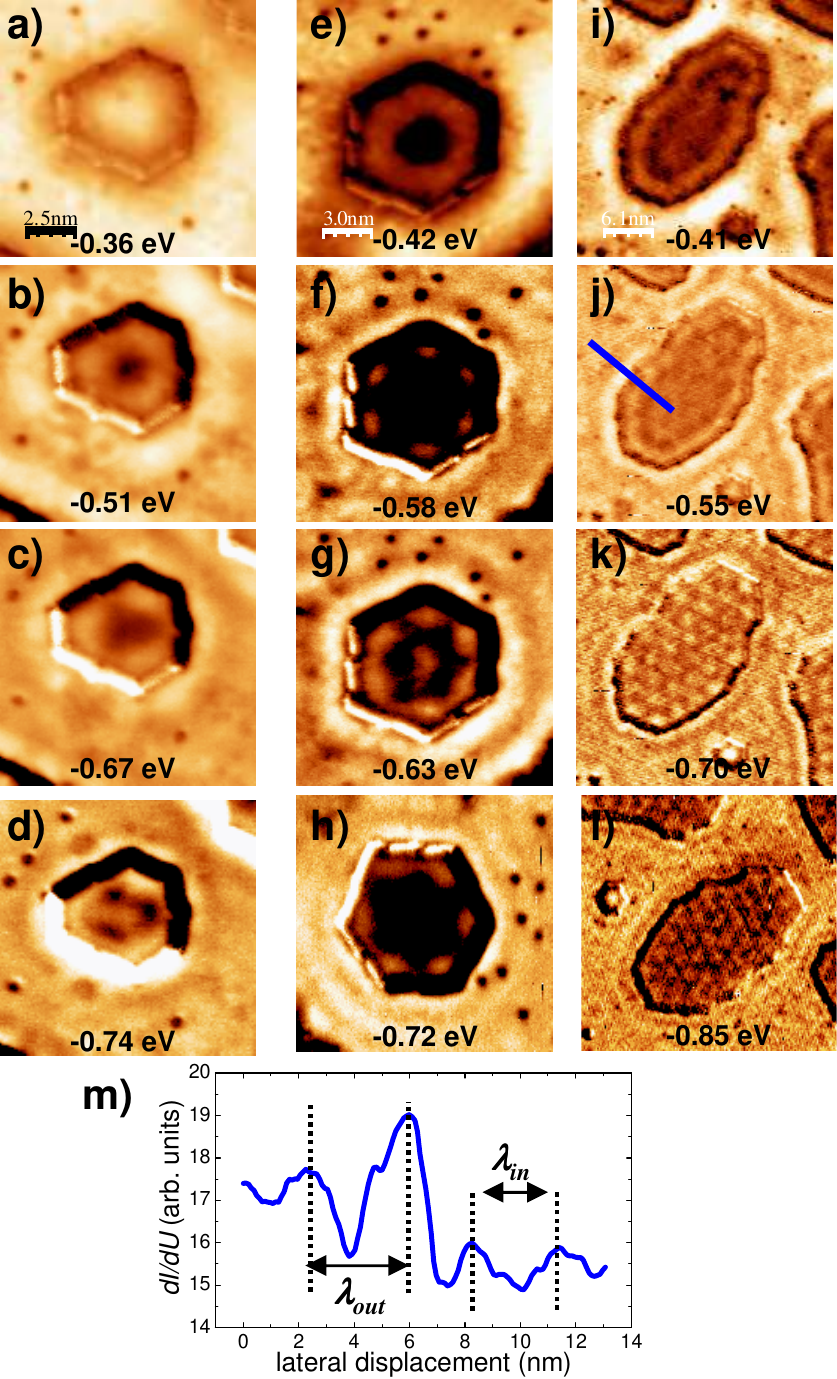}
\caption{\label{stw} (a)-(l) $dI/dU$ maps of three different
  graphene QDs  recorded at the energies $E=U\cdot e$ marked;
  $U_{\rm mod} =10$ mV; (a)-(d),(i)-(l) $I=0.2$\,nA, (e)-(h) $I=0.5$ nA; all images exhibit a standing wave around the island; wave length decreases with
  decreasing energy; blue line in (j) marks the line for the line scan shown in (m); (m) line scan along the line shown in (j); the deduced wave lengths of the standing waves inside ($\lambda_{\rm in}$) and outside ($\lambda_{\rm out}$) the QD are marked.}
\end{figure}

\begin{figure}[h!]
\includegraphics[width=8.5cm]{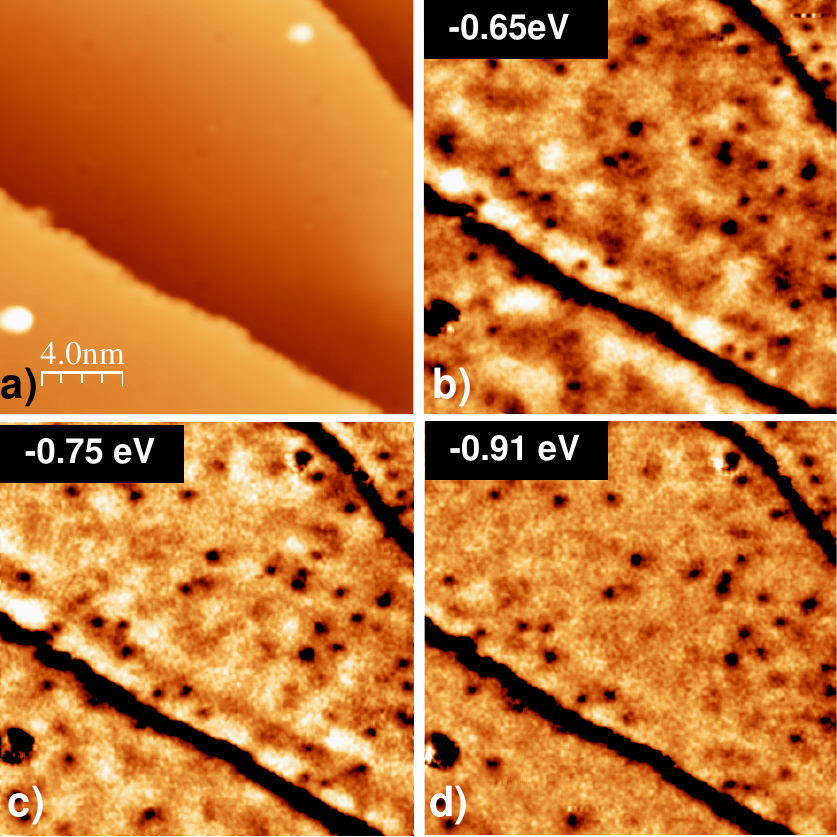}
\caption{\label{stwIr} (a) STM image of the Ir(111) surface with two step edges;
$U=-0.2$ V; $I=0.3$ nA; (b)-(d) $dI/dU$ maps of the same area recorded at the energies $E=U\cdot e$ marked ($I=0.3$\,nA,
  $U_{\rm mod} =10$ mV); standing waves are visible at the upper and the lower side of the step edges.}
\end{figure}

Figure \ref{stw} shows $dI/dU$ maps of three different quantum dots where the contrast is
tuned in order to see the standing wave outside the QD. Obviously the wavelengths of these standing waves
$\lambda_{\rm out}$ is decreasing with decreasing energy. By line scans,
as shown in Fig. \ref{stw}m for the $dI/dU$ map in Fig. \ref{stw}j, we deduced the $E(\pi/\lambda_{\rm out})$
dispersion shown in Fig. 4g of the main text. Since the measured standing wave pattern exhibits
half the wave length of the impinging Bloch wave, $\Delta k= \pi/\lambda_{\rm out}$ is used such that the
dispersion $E(\Delta k)$ can directly be compared with data from angular resolved photoelectron spectroscopy (ARPES) \cite{Rader}.\\
The standing waves are also observed at step edges of the Iridium(111) surface not covered by graphene.
Figure \ref{stwIr}a shows an STM image of the uncovered Ir(111) surface with two step edges. The two white dots are larger
adsorbates on the surface. The $dI/dU$ images exhibit standing waves at the step edges whose wave length decreases with decreasing energy. In addition, remaining oxygen adsorbates on the surface are visible as black dots which induce an additional complicated scattering pattern on the terraces. The wave lengths $\lambda_{\rm Ir}$ of the standing waves at the step edges are determined by line scans averaging along the step edge and the resulting $E(\pi/\lambda_{\rm Ir})$ is also plotted in Fig. 4g of the main text. The symbols in Fig. 4g of the main text exhibit  a very similar steepness of $E(\pi/\lambda_{x})$ ($x={\rm in}, {\rm out}, {\rm Ir}$) for all three measurements, but the absolute values are
lowest for the pure Ir(111) surface, slightly higher (about 50 meV) for the standing waves around the graphene QDs and the highest
for the standing waves inside the graphene QDs being another 100 meV higher. The photoemission data \cite{Rader} show the same
energy shift of the dispersion of the state S$_0$ by about 150 meV between uncovered Ir(111) and Ir(111) completely covered with graphene.

\newpage
\section{Estimate of confined energies}

As described in the main text, we estimate the energy of confined states by the zeros of the
  first two Bessel functions according to:

\begin{equation}\label{Eq:Bessels}
J_n(k_n\cdot r) = 0,\quad n=0,1.
\end{equation}

with eigenenergies $E_n=\hbar v_D k_n$, a Dirac velocity of graphene
$v_D=1\cdot 10^6$ m/s and Radius $r$ of the island.  The first two
peak energies $E_0$ and $E_1$ of the experiment are determined with
respect to the Dirac point $E_D$ for different islands. Since
  the shapes of larger islands become more irregular (see
  Fig. \ref{overview}), only the nine smallest islands are
  considered. The average island radius $r$ is deduced from the
island area $A$ by $r=\sqrt{A/\pi}$. The resulting $k_n\cdot r$ is
plotted as a function of $r$ for the two peak energies closest to
$E_D$ in Fig. 3 (h) of the main text. For a few of the islands, the
  energy of the 2$^{\rm nd}$ resonance in the $dI/dU$ curve is not
  well defined due to broadening of lineshapes by finite state
  lifetime, and thus not considered in the present
  analysis. For the smallest island, only the first resonance energy
  lies within the Ir projected band gap. Obviously, reasonable agreement of the model with the
experiment is found. The discrepancy increases with radius which is attributed
to the more non-circular shape of the larger islands.